\documentclass[aps,aps,prl,twocolumn,floatfix,showpacs]{revtex4}
\usepackage{amsmath,amssymb,graphicx,bm}
\begin{document}
\title{A mean-field theory of Anderson localization}

\author{V.  Jani\v{s}} \author{ J. Koloren\v{c}}

\affiliation{Institute of Physics, Academy of Sciences of the Czech
  Republic, Na Slovance 2, CZ-18221 Praha 8, Czech Republic}
\email{janis@fzu.cz, kolorenc@fzu.cz}

\date{\today}


\begin{abstract}
Anderson model of noninteracting disordered electrons is studied in high
spatial dimensions. We find that off-diagonal one- and two-particle
propagators behave as gaussian random variables w.r.t.  momentum
summations. With this simplification and with the electron-hole symmetry
we reduce the parquet equations for two-particle irreducible vertices to a
single algebraic equation for a local vertex. We find a disorder-driven
bifurcation point in this equation signalling vanishing of diffusion and
onset of Anderson localization. There is no bifurcation in $d=1,2$ where
all states are localized. A natural order parameter for Anderson
localization pops up in the construction.
\end{abstract}

\maketitle 
Understanding mobility of (free) particles in random media has been a
challenging theoretical problem for many decades. It became clear from the
early days of the study of systems with randomly distributed scatterers
that the particle movement has a diffusive character described at long
distances by a diffusion equation. A breakthrough in the conceptual
perception of random systems was achieved in Ref.~\cite{Anderson58}.
P. W. Anderson demonstrated there on a simple model that for a
sufficiently strong disorder the particle remains confined in a finite
volume and fails to diffuse to long distances. Since then, the
disorder-induced absence of diffusion, called Anderson localization, has
attracted a lot of attention of condensed matter theorists. In spite of
years of intensive studies, the phenomenon of Anderson localization has
not yet been fully understood. It is mostly due to many facets of this
difficult problem.

There are two complementary tools, numerical and analytical, to tackle
Anderson localization and the metal-insulator transition. Neither of them
is, however, able to answer all questions about the disorder-induced
vanishing of diffusion. While the former deals with finite lattices and
many configurations of the random potential \cite{Kramer93}, the latter
deals mostly with the thermodynamic limit and configurationally averaged
quantities \cite{Lee85}. Although the existence of Anderson localization
was proved rigorously to exist in any finite dimension if the disorder is
sufficiently strong \cite{Frohlich83}, both numerical and analytical
approaches work preferably in rather low dimensions ($d=2 + \epsilon$). As
a consequence, a standard mean-field (high-dimensional) theory of
Anderson localization is missing, except for special solutions
on a Bethe lattice \cite{Efetov97}. Even though a self-consistent,
mean-field-type theory of Anderson localization was formulated
\cite{Vollhardt80b}, it was derived only within the weak-scattering,
low-dimensional limit. A systematic mean-field theory should come out
of the asymptotic limit to high spatial dimensions \cite{Janis99a}.
 
The aim of this Letter is to employ the limit to high spatial dimensions
for developing a mean-field theory for the Anderson localization
transition. We use the parquet approach summing up systematically nonlocal
vertex corrections to the mean-field, $d=\infty$, solution
\cite{Janis01b}. We show how the parquet equations for the irreducible
two-particle vertices can be simplified and solved in the asymptotic limit
to high spatial dimensions. We use the asymptotic solution of the parquet
equations to build up a quantitative theory of the disorder-driven
metal-insulator transition with a mean-field critical behavior, i.~e.,
independent of the spatial dimension.
 
The existence or nonexistence of diffusion can be determined from the
electron-hole correlation function defined from the two-particle resolvent
as   $ \Phi(z_1,z_2;{\bf q}) = N^{-2} \sum_{{\bf k}{\bf k}'} G^{(2)}_{{\bf
k}{\bf k}'}(z_1,z_2;{\bf q})$, where $z_1$ and $z_2$ are complex energies.
A specific element of this function with energies $z_1=E_F + \omega + i0^+$
and $z_2= E_F - i0^+$, denoted $\Phi^{AR}_{E_F}(\mathbf{q},\omega)$, is
used to determine the diffusion constant \cite{Vollhardt80b,Janis03a}
\begin{equation}
  \label{eq:AR-diffusion}
  n_F D = \lim\limits_{\omega\to0 }\frac{\omega^2}{4\pi} \nabla^2_{q}
  \Phi^{AR}_{E_F}(\mathbf{q},\omega)\big|_{q=0}
\end{equation}
where $n_F$ is the density of states at the Fermi energy $E_F$. Vanishing
of the diffusion constant $D$ indicates the absence of diffusion in
the system.

In systematic theories we do not approximate directly either the Green
function $G^{(2)}$ or the correlation function $\Phi^{AR}$, but rather the
two-particle vertex $\Gamma$ defined from \cite{Note1}
\begin{multline*}
G^{(2)}_{{\bf k}{\bf k}'}({\bf q}) = G_+({\bf k}) G_-({\bf k} + \mathbf{q})
\left[\delta({\bf k} - {\bf k}')\right. \\ \left. + \Gamma_{{\bf k}{\bf
k}'}({\bf q}) G_+({\bf k}') G_-({\bf k}' + \mathbf{q})\right] \
\end{multline*}
where $G_\pm(\mathbf{k}) \equiv G(\mathbf{k},z_\pm)$ are averaged
one-particle resolvents.

The simplest theory for strongly disordered systems is the
coherent-potential approximation (CPA), being an exact solution in
$d=\infty$. Only the diagonal part of the one-electron resolvent $G(z) =
N^{-1} \sum_{\mathbf{k}} G(\mathbf{k},z)$ is relevant in the CPA, since it
completely neglects coherence between spatially distinct scatterings. It is
a consistent mean-field theory only for one-electron functions. Nonlocal
parts of two-particle functions in high dimensions do not vanish. The
off-diagonal elements of the one-electron propagators cannot be neglected
and have to be taken expliciltly into consideration \cite{Janis99a}.
Further on, the CPA, due to its local character, is degenerate and cannot
distinguish between scatterings of electrons and holes. To be able to
resolve various types of two-particle scatterings in noninteracting
systems we have to go beyond the mean-field, local approximation.

A systematic (diagrammatic) expansion around the $d=\infty$ limit can be
performed by using the off-diagonal one-electron CPA resolvent
$\bar{G}(\mathbf{k},z) \equiv G(\mathbf{k},z) - G(z)$. Three
nonequivalent Bethe-Salpeter equations with nonlocal propagators can be
constructed for the full vertex $\Gamma$ \cite{Janis01b}. Here we will use
only the Bethe-Salpeter equations from the electron-hole and the
electron-electron scattering channels \cite{Note2}. They can be
represented as
\begin{subequations}\label{eq:BS}\begin{align}
\begin{split}\label{eq:BS-eh}
\Gamma_{\mathbf{k}\mathbf{k}'}(\mathbf{q})& =
\bar{\Lambda}^{eh}_{\mathbf{k}\mathbf{k}'}(\mathbf{q}) + \frac
1N\sum_{\mathbf{k}''}
\bar{\Lambda}^{eh}_{\mathbf{k}\mathbf{k}''}(\mathbf{q}) \\ &\quad \times
\bar{G}_+(\mathbf{k}'') \bar{G}_-(\mathbf{k}'' + \mathbf{q})
\Gamma_{\mathbf{k}''\mathbf{k}'}(\mathbf{q}) \ ,\end{split}\\
\begin{split}\label{eq:BS-ee}
\Gamma_{\mathbf{k}\mathbf{k}'}(\mathbf{q}) &=
\bar{\Lambda}^{ee}_{\mathbf{k}\mathbf{k}'}(\mathbf{q}) + \frac
1N\sum_{\mathbf{k}''}
\bar{\Lambda}^{ee}_{\mathbf{k}\mathbf{k}''}(\mathbf{q} + \mathbf{k}'  -
\mathbf{k}'')\\ &\quad  \times \bar{G}_+(\mathbf{k}'') \bar{G}_-(\mathbf{Q}
- \mathbf{k}'') \Gamma_{\mathbf{k}''\mathbf{k}'}(\mathbf{q}+ \mathbf{k} -
\mathbf{k}'')\ ,
\end{split}
\end{align}\end{subequations}
respectively. We used bar in the irreducible vertices $\bar{\Lambda}^{eh}$
and $\bar{\Lambda}^{ee}$ to indicate that the Bethe-Salpeter equations are
constructed with the off-diagonal resolvents only. Hence, in the
infinite-dimensional case $\bar{G}_\pm(\mathbf{k})=0$ and
$\bar{\Lambda}^{eh,ee}=\gamma$, where $\gamma$ is the full local CPA
vertex \cite{Janis01b}. For simplicity of notation we denoted $\mathbf{Q}
\equiv \mathbf{q} + \mathbf{k} + \mathbf{k}'$. Notice that $\mathbf{q}$ is
the momentum conserved for scatterings in the electron-hole channel,
Eq.~\eqref{eq:BS-eh}, while $\mathbf{Q}$ is conserved in the
electron-electron channel, Eq.~\eqref{eq:BS-ee}.
 
To reach the strong-disorder limit with a diffusionless regime we have to
determine the irreducible vertices $\bar{\Lambda}^{eh}$ and
$\bar{\Lambda}^{ee}$ self-consistently. The parquet construction
provides a suitable framework for this purpose. It is based on the
observation that \textit{reducible diagrams} in one channel are
\textit{irreducible} in the other, topologically distinct channels. If we
approximate the vertex irreducible in all channels by the local CPA vertex
$\gamma$, take into account only the $eh$ and $ee$ channels, and realize
that the full vertex is a sum of reducible and irreducible vertices
in any channel, we end up with a fundamental parquet equation
\begin{equation}\label{eq:parquet-equation}
\Gamma_{\mathbf{k}\mathbf{k}'}(\mathbf{q}) =
\bar{\Lambda}^{eh}_{\mathbf{k}\mathbf{k}'}(\mathbf{q}) +
\bar{\Lambda}^{ee}_{\mathbf{k}\mathbf{k}'}(\mathbf{q}) - \gamma\ .
\end{equation}
The minus sign at the CPA vertex compensates for the identical local part
in both  $\bar{\Lambda}^{eh}$ and $\bar{\Lambda}^{ee}$. A couple of
(nonlinear) parquet equations determining the irreducible vertices
$\bar{\Lambda}^{eh}$ and $\bar{\Lambda}^{ee}$ as functions of $\gamma$ and
$\bar{G}_\pm$ are obtained by replacing the full vertex $\Gamma$ in
Eqs.~\eqref{eq:BS} by Eq.~\eqref{eq:parquet-equation}.

Prior to solving the parquet equations for $\bar{\Lambda}^{eh}$ and
$\bar{\Lambda}^{ee}$ we utilize the electron-hole symmetry expressed as an
identity for two-particle vertices
$\Gamma_{\mathbf{k}\mathbf{k}'}(\mathbf{q}) =
\Gamma_{\mathbf{k}\mathbf{k}'}(- \mathbf{q} - \mathbf{k} - \mathbf{k}')$
and $\bar{\Lambda}^{ee}_{\mathbf{k}\mathbf{k}'}(\mathbf{q}) =
\bar{\Lambda}^{eh}_{\mathbf{k}\mathbf{k}'}(-\mathbf{q}- \mathbf{k} -
\mathbf{k}')$. The electron-hole transformation maps Eq.~\eqref{eq:BS-eh}
onto Eq.~\eqref{eq:BS-ee}. The two-particle electron-hole symmetry is a
consequence of the time-reversal invariance  the one-electron resolvent
$\bar{G}(\mathbf{k},z) = \bar{G}(-\mathbf{k},z)$ used in the
Bethe-Salpeter equations \eqref{eq:BS}. This symmetry
actually reduces the number of parquet equations to a single nonlinear
integral equation for $\bar{\Lambda}_{\mathbf{k}\mathbf{k}'}(\mathbf{q})
\equiv \bar{\Lambda}^{ee}_{\mathbf{k}\mathbf{k}'}(\mathbf{q}) =
\bar{\Lambda}^{eh}_{\mathbf{k}\mathbf{k}'}(-\mathbf{q}- \mathbf{k} -
\mathbf{k}')$.

Generally, the parquet equations are unsolvable due to momentum
convolutions in the Bethe-Salpeter equations. The limit to high spatial
dimensions leads to suppression of spatial fluctuations resulting in
simplifications of momentum convolutions \cite{Janis01a}. We take
advantage of these simplifications. We start with the leading asymptotic
term in the off-diagonal propagator $\bar{G}(\mathbf{k},z)$ that on a
$d$-dimensional hypercubic lattice with the hopping amplitude $t$ reads
\begin{multline}\label{eq:1P-nonlocal}
\bar{G}(\mathbf{k},z) \doteq
\frac{t}{\sqrt{d}}\sum_{\nu=1}^d \cos(k_\nu) \int d\epsilon\rho(\epsilon)
G(z-\Sigma(z)-\epsilon)^2\\  = tx(\mathbf{k})\langle G(z)^2\rangle\ ,
\end{multline}
where $\rho$ is the density of states and $\Sigma$ the local (CPA)
self-energy. We replace the off-diagonal one-electron propagators in the
parquet equations with this asymptotic representation.

The simplest and most important convolution is a two-particle bubble
$\bar{\chi}(\mathbf{q}) = N^{-1}\sum_{\mathbf{k}} \bar{G}_+(\mathbf{k})
\bar{G}_-(\mathbf{k} + \mathbf{q})$. Its asymptotic behavior can be found
from the following formula
\begin{subequations}\label{eq:convolutions}
\begin{equation} \label{eq:convolution-ff}
\frac 1N\sum_{\mathbf{k}}x(\mathbf{k})x(\mathbf{k}+\mathbf{q}) = \frac 12
X(\mathbf{q}) = \frac 1{2d} \sum_{\nu=1}^d \cos(q_\nu)
\end{equation}
where we denoted $X(\mathbf{q})$ a two-particle (bosonic) dispersion
function. Other possible convolutions of the generic fermionic
and bosonic dispersion functions are
\begin{align}\label{eq:convolution-fb}
\frac 1N\sum_{\mathbf{q}'}X(\mathbf{q}' +
\mathbf{q})x(\mathbf{q}'+\mathbf{k}) &= \frac 1{2d} x(\mathbf{q}-
\mathbf{k})\ ,
\\ \label{eq:convolution-bb}
\frac 1N\sum_{\mathbf{q}'}X(\mathbf{q}' + \mathbf{q}_1)X(\mathbf{q}'
+\mathbf{q}_2) &= \frac 1{2d} X(\mathbf{q}_1- \mathbf{q}_2)\ .
\end{align}
\end{subequations}
We can see that the fermionic and bosonic dispersion functions form a
closed algebra with respect to momentum summations. The
elementary convolutions \eqref{eq:convolutions} manifest the generation of
the factor $d^{-1}$ due to mixing of two-particle propagations from
different scattering channels. The dispersion functions then behave in the
leading asymptotic order of $d^{-1}$ as \textit{gaussian random variables}
when momentum summations are performed.

To make the calculations in high spatial dimensions more mean-field-like,
we replace the bare fermionic and bosonic dispersion functions with the
respective off-diagonal one- and two-particle propagators. That is, we use
$\bar{G}$ instead of $x$ and $\bar{\chi}$ instead of $X$. These quantities
are directly proportional in the leading asymptotic order.  Without loosing
the asymptotic accuracy we can extend
relations~\eqref{eq:convolutions} to genuine mean-field expressions
$N^{-1}\sum_{\mathbf{q}'} \bar{\chi}(\mathbf{q}' + \mathbf{q})
\bar{G}_\pm(\mathbf{q}' + \mathbf{k}) = W\bar{G}_\pm(\mathbf{q} -
\mathbf{k})/4d$,  $N^{-1}\sum_{\mathbf{q}} \bar{\chi}(\mathbf{q} +
\mathbf{q}_1) \bar{\chi}(\mathbf{q} + \mathbf{q}_2) =
W\bar{\chi}(\mathbf{q}_1 - \mathbf{q}_2)/4d$, where we used
$W=t^2\langle G_+^2\rangle\langle G_-^2\rangle$.

It is evident from Eq.~\eqref{eq:1P-nonlocal} and
Eq.~\eqref{eq:convolutions} that the two-particle vertices can be
represented as functions of the generic off-diagonal fermionic
$\bar{G}(\mathbf{k})$ and bosonic $\bar{\chi}(\mathbf{q})$ functions. To
find the leading high-dimensional asymptotics of the solution of the
parquet equation for $\bar{\Lambda}_{\mathbf{k}\mathbf{k}'}(\mathbf{q})$
we keep only the leading $d^{-1}$ terms for each specific momentum
dependence of the vertex. It is easy to demonstrate that the parquet
equation then simplifies in the leading asymptotic limit $d\to\infty$ to
an algebraic equation
\begin{subequations}\label{eq:parquet-simplified}
\begin{equation}\label{eq:parquet-momentum}
\bar{\Lambda}_{\mathbf{k}\mathbf{k}'}(\mathbf{q}) = \bar{\Lambda}
(\mathbf{q}) = \gamma + \bar{\Lambda}_0 \frac {\bar{\Lambda}_0 \bar{\chi}
(\mathbf{q})} {1 - \bar{\Lambda}_0\bar{\chi}(\mathbf{q})}
\end{equation}
where $\bar{\Lambda}_0 = N^{-1}\sum_{\mathbf{q}}\bar{\Lambda}(\mathbf{q})$.
The high-dimensional irreducible two-particle vertex is completely
determined from a single local parameter $\bar{\Lambda}_0$ and the
two-particle bubble $\bar{\chi}(\mathbf{q})$. Summing both sides of
Eq.~\eqref{eq:parquet-momentum} over momenta we obtain
\begin{equation}\label{eq:parquet-local}
\bar{\Lambda}_0 = \gamma + \bar{\Lambda}_0 \frac 1N \sum_{\mathbf{q}} \frac
{\bar{\Lambda}_0 \bar{\chi}(\mathbf{q})}{1 - \bar{\Lambda}_0
\bar{\chi}(\mathbf{q})} \ .
\end{equation}\end{subequations}

Equations~\eqref{eq:parquet-simplified} were derived from the leading
high-dimensional asymptotics, but can be used in any finite dimension. It
is, however, mandatory that the proper $d$-dimensional momentum summation
is used on a $d$-dimensional lattice. We cannot directly use the gaussian
rules to evaluate the summation over momenta in $d$ dimensions as we did
during the derivation of these equations. The integrand  would be singular
(nonintegrable) in the gaussian evaluation. To assess the asymptotic
high-dimensional behavior of the two-particle irreducible  vertex we have
to realize that in deriving Eq.~\eqref{eq:parquet-simplified} each
independent one-dimensional momentum integration over the components $q_i$
with $i=1, 2,\ldots, d$ can contain maximally squares of the dispersion to
remain within the leading asymptotics. Hence, on a $d$-dimensional lattice
we can build maximally $d$ pairs of the two-particle bubbles $\bar{\chi}$.
The integrand in Eq.~\eqref{eq:parquet-local} therefore collapses to a
polynomial of order $d$. Using the gaussian integration rules we
explicitly obtain
\begin{equation*}
f_d(a) =
\frac 1N \sum_{\mathbf{q}} \frac {\bar{\Lambda}_0 \bar{\chi}(\mathbf{q})}{1
- \bar{\Lambda}_0 \bar{\chi}(\mathbf{q})}\bigg|_d \equiv \sum_{n=1}^{d}
\frac {(2n)!}{(2d)^n n! }\ \left(a\right)^n
\end{equation*}
where we denoted $a = W^2\bar{\Lambda}_0^2/8$. The asymptotic limit of the
sum $f_{d\to\infty}(a)$ converges for $a<1/2$, i.~e., for
$\bar{\Lambda}_0^2 < 4/W^2$. The critical value of the randomness
$\gamma^2 = 4/W^2$ defines an ultimate upper bound beyond which
perturbation theory around $d=\infty$ in powers of $d^{-1}$ does not
converge and becomes nonanalytic. In realistic models, however, such an
extreme value cannot be reached, apart from tiny regions around band edges
and in satellite bands.

The one-electron functions $G,\Sigma$ and the local two-particle vertex
$\gamma$ entering Eqs.~\eqref{eq:parquet-simplified} were assumed to be
taken from the CPA. There we have $\gamma = \lambda/(1 - \lambda G_+ G_-)$
and $\lambda= (\Sigma_+ - \Sigma_-)/(G_+ - G_-)$. If we analogously define
$\bar{\Lambda}_0 = \Lambda_0/(1 - \Lambda_0G_+ G_-)$ and $\chi(\mathbf{q})
= \bar{\chi}(\mathbf{q}) + G_+ G_-$ we can represent the asymptotic form of
the full two-particle vertex as follows
\begin{multline}\label{eq:vertex-full}
\Gamma_{\mathbf{k}\mathbf{k}'}(\mathbf{q}) = \gamma \\ + \Lambda_0
\left[\frac{\bar{\Lambda}_0\bar{\chi}(\mathbf{q})} {1 - \Lambda_0
\chi(\mathbf{q})} + \frac{\bar{\Lambda}_0\bar{\chi}(\mathbf{k} +
\mathbf{k}' + \mathbf{q})} {1 - \Lambda_0 \chi(\mathbf{k} + \mathbf{k}' +
\mathbf{q})}\right]\ .
\end{multline}
The full nonlocal CPA vertex can be recovered from the above expression if
we put $\bar{\Lambda}_0=\gamma$ and neglect the last term on the r.h.s. of
Eq.~\eqref{eq:vertex-full}. The term neglected in the CPA, however,
restores the electron-hole symmetry in the asymptotic vertex in high
dimensions.

Up to now we have analyzed the two-particle asymptotics with the
one-electron propagators fixed by the CPA. To reproduce diffusion in this
approach we have to match correctly the irreducible vertex calculated from
the parquet equations \eqref{eq:parquet-simplified} and the one-electron
self-energy. We hence have to go beyond the CPA even in the one-electron
propagators. To do so consistently we use the Ward identity and determine
the imaginary part of the self-energy from the two-particle irreducible
vertex via
\begin{equation}\label{eq:self-energy}
\Im\Sigma(E+i0^+) = \Lambda_0(E+i0^+, E-i0^+) \Im G(E+i0^+) \ .
\end{equation}
The real part of the self-energy is determined from the Kramers-Kronig
relation \cite{Janis01b}. Equation~\eqref{eq:self-energy} completes the
parquet equation~\eqref{eq:parquet-local}. Both equations together with
the Kramers-Kronig relation have to be solved simultaneously to achieve
full self-consistence between $\Lambda_0$ and $G$ calculated from $\Sigma$
via the Dyson equation.

The redefinition of the self-energy in Eq.~\eqref{eq:self-energy} is
important, since only with it we recover the diffusion pole in the vertex
functions. That is, we obtain $\Lambda_0(E+i0^+,
E-i0^+)\chi(\mathbf{0})=1$, whenever the parquet equation
\eqref{eq:parquet-local} allows for a positive solution. The existence of
the diffusion pole is essential for the diffusion constant from
Eq.~\eqref{eq:AR-diffusion} to be non-zero (positive). We can immediately
conclude from simple power counting in the momentum integral of
Eq.~\eqref{eq:parquet-local} that there is no positive solution for
$\bar{\Lambda}_0(E+i0^+, E-i0^+)$ in low dimensions, $d=1,2$, since the
diffusion pole would be nonintegrable. Consequently, no diffusion pole can
exist and the diffusion constant from Eq.~\eqref{eq:AR-diffusion} vanishes
in $d=1,2$.

In higher dimensions we can expand the r.h.s. of
Eq.~\eqref{eq:parquet-local} in powers of the local vertex
$\bar{\Lambda}_0$. We then obtain a mean-field-like cubic equation
\begin{equation}\label{eq:parquet-high_dim}
\bar{\Lambda}_0 = \gamma + C_d\bar{\Lambda}_0^3
\end{equation}
with $C_d = \lim_{\bar{\Lambda}_0\to0}\
N^{-1}\sum_{\mathbf{q}}\bar{\chi}(\mathbf{q})^2/(1 - \bar{\Lambda}_0
\bar{\chi}(\mathbf{q}))^2$. This constant is generally a decreasing
function of the spatial dimension $d$ and approaches zero in the limit
$d\to\infty$ via $C_d\sim W^2/8d$. Due to the existence of the diffusion
pole in Eq.~\eqref{eq:parquet-local} the constant $C_d$ becomes infinite
in $d\le 4$. Equation~\eqref{eq:parquet-high_dim} derived from a Taylor
expansion in the local vertex $\bar{\Lambda}_0$ does not survive in this
form to low dimensions.

\begin{figure}
  \resizebox{8.5cm}{!}
{\includegraphics{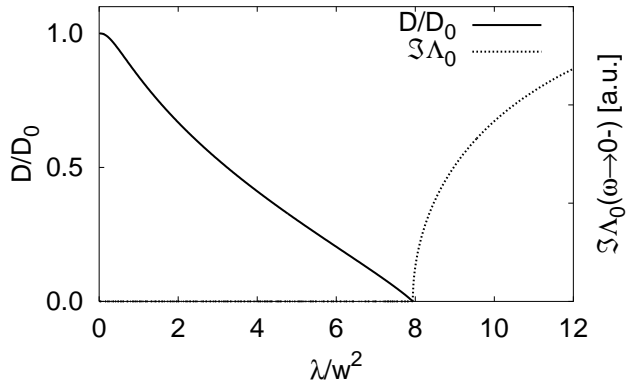}}
\caption{\label{fig:Born-order}Diffusion constant $D$ and the order
parameter in the localized phase $\Im\Lambda_0$ calculated from
Eq.~\eqref{eq:parquet-high_dim}. We used a semielliptic energy band with
the bandwidth $2w$, the self-consistent Born approximation for the
self-energy, and set $C_d=0.1W^2$.}
\end{figure}

\begin{figure}
  \resizebox{8.5cm}{!}
{\includegraphics{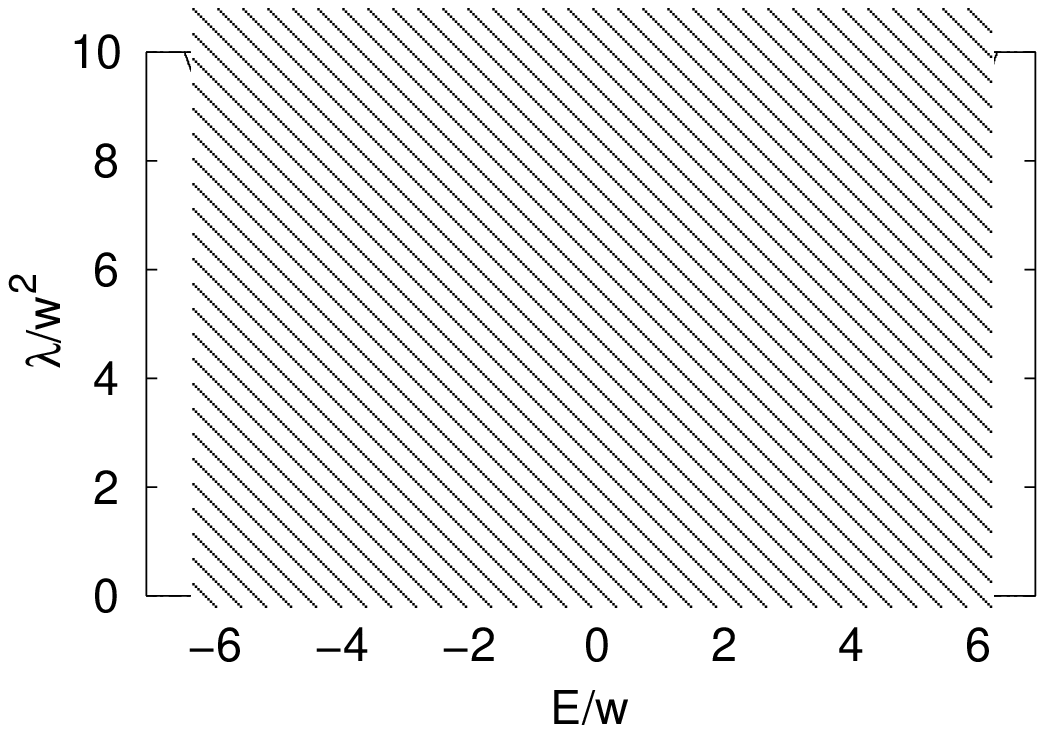}}
\caption{\label{fig:Born-edges} Phase diagram for the same setting as in
Fig.~\ref{fig:Born-order}. The hatched area denotes localized states.}
\end{figure}

Equation \eqref{eq:parquet-high_dim} has generally three solutions for
$\bar{\Lambda}_0(E+i0^+, E-i0^+)$. For sufficiently small disorder
strengths, $\gamma < \gamma_c$, all three solutions are real. A
perturbative solution is of order $\gamma$, while two nonperturbative
solutions are of order $\pm\sqrt{1/C_d}$. The perturbative solution
increases and the module of the nonperturbative ones decreases with
increasing the disorder strength. At a critical randomness
$3C_d\bar{\Lambda}_0^2 = 1$, or equivalently $\gamma_c = \sqrt{4/27C_d}$,
the two positive solutions merge and move into the complex plane for
$\gamma > \gamma_c$. Disappearance of positive solutions for
$\bar{\Lambda}_0(E+i0^+, E-i0^+)$ leads to suppression of the diffusion
pole and simultaneously to vanishing of the diffusion constant. Quantity
$\Im\bar{\Lambda}_0(E+i0^+, E-i0^+)$, emerging beyond the critical point
in the localized phase ($\gamma >\gamma_c$),  plays the role of an order
parameter for Anderson localization, see Fig.~\ref{fig:Born-order}. A
typical phase diagram for localized-extended states calculated from
Eq.~\eqref{eq:parquet-high_dim} is plotted in Fig.~\ref{fig:Born-edges}.
Although the mean-field equation \eqref{eq:parquet-high_dim} does not
predict the precise position of the mobility edges, it determines the
mean-field universal properties accurately.

To conclude, we derived a mean-field approximation for two-particle
irreducible  vertices motivated and justified by the asymptotic limit to
high dimensions. We succeeded in deriving an algebraic equation for the
local irreducible vertex with a bifurcation point at which the diffusion
constant vanishes and a real irreducible vertex becomes complex. A fully
consistent and controllable mean-field-like theory of the disorder-driven
vanishing of diffusion and Anderson localization was thereby achieved. It
correctly reproduces the low and high-dimensional limits and allows for
further systematic improvements.

Research on this problem was carried out within a project AVOZ1-010-914
of the Academy of Sciences of the Czech Republic and supported in part by
Grant No. 202/01/0764 of the Grant Agency of the Czech Republic.


\begin{thebibliography}{99}
\bibitem{Anderson58} P.~W.~Anderson, \newblock Phys. Rev. {\bf109}, 1492
(1958).
 
\bibitem{Kramer93} B.~Kramer and A.~MacKinnon, \newblock Rep. Mod. Phys.
  {\bf 56}, 1472 (1993).

\bibitem{Lee85} P.~A.~Lee and R.~V.~Ramakrishnan, \newblock Rev. Mod. Phys.
  {\bf 57}, 287 (1985).

\bibitem{Frohlich83} J. Fr\"ohlich and T. Spencer, \newblock Commun. Math.
Phys. \textbf{88}, 151 (1983).

\bibitem{Efetov97} K. Efetov, \textit{Supersymmetry in disorder and chaos},
\newblock Cambridge University Press, Cambridge 1997.

\bibitem{Vollhardt80b} D.~Vollhardt and P. W\"olfle, \newblock Phys. Rev.
B{\bf 22}, 4666 (1980) and D.~Vollhardt and P.~W\"olfle, \newblock in
\textit{Electronic Phase Transitions}, W.~Hanke and Yu.~V.~Kopaev (eds),
(Elsevier Science Publishers B. V., Amsterdam 1992)..

\bibitem{Janis99a} V.~Jani\v s, \newblock Phys. Rev. Lett. \textbf{83},
2781 (1999).

\bibitem{Janis01b} V.~Jani\v s, \newblock Phys. Rev. B{\bf 64}, 115115
  (2001).

\bibitem{Janis03a} V.~Jani\v s, J.~Koloren\v c, and V. \v Spi\v cka,
\newblock Eur. Phys. J. B{\bf 35}, 77 (2003).

\bibitem{Note1} Whenever possible we suppress the energy variables in the
Green functions in order to keep the notation simple. Energies are not
dynamical variables and are conserved in elastic scatterings of
noninteracting particles. The energy variables are hence easily
identifiable in all expressions.
 
\bibitem{Note2} The third two-particle scattering
channel consists of one-electron self-corrections and is called
vertical channel (cf. Ref.~\cite{Janis01b}).

 
\bibitem{Janis01a} V.~Jani\v s and D. Vollhardt, \newblock Phys. Rev. B{\bf
63}, 125112 (2001).


\end{thebibliography}
\end{document}